\documentclass[a4paper]{article}

\usepackage[latin1]{inputenc}
\usepackage[english]{babel}
\usepackage[final]{graphicx}
\usepackage{verbatim}

\usepackage{amsmath,amsfonts,amssymb,amsthm}
\usepackage{latexsym}

\newcommand{\structure}[1]{$^{#1}$}

\newcommand{\authorstructure}[2][]{$ $\hspace{-5mm}$^{#1}$ {\small \it #2}}

\newenvironment{Aabstract}
	{\vspace{5mm}
	 \begin{center}\bf Abstract\end{center}
	 \begin{quote}\small}
	{\end{quote}\vspace{5mm}}

\begin{document}

\begin{center}
{\LARGE Sensitivity of Gamma Astroparticle Experiments to the Detection of \mbox{Photon Oscillations}}
\\[5mm]
{\sc Alessandro De Angelis\structure{a,b}, 
             Oriana Mansutti\structure{a},
             Reynald Pain\structure{c}}
\end{center}

\authorstructure[a]{Dipartimento di Fisica, 
                   Universit\`a di Udine, 
                   via delle Scienze~208, 33100~Udine, Italy}

\authorstructure[b]{INFN, \,Sezione \,di \,Trieste, \,Gruppo \,Collegato \,di \,Udine, \,via \,delle \,Scienze \,208, \,33100~\,Udine, \,Italy}
                   
\authorstructure[c]{LPNHE, Universit\'es Paris VI \& VII and IN2P3/CNRS, Paris, France}

\begin{Aabstract}
The mixing of the photon with a hypothetical sterile paraphotonic state 
would have consequences on the cosmological propagation of photons.  
Observations of gamma-rays from active galactic nuclei in GLAST and MAGIC will open a new domain in the search
for such a phenomenon. 
\end{Aabstract}

The existence of a second photon (paraphoton) mixing to the ordinary one 
was first postulated in \cite{geo} to explain a presumed anomaly in the 
spectrum of the Cosmic Microwave Background (CMB). 
In that model, the anomaly was 
attributed to a mass mixing of the two photons analogous to the
oscillation of neutrinos. 
An ordinary photon oscillates with the time $t$ in such a way that its 
probability
to stay as such can be written as
\begin{equation}
P(t) = 1-\sin^2(2\phi)\,\sin^2{(\rho \mu^2 \, t/\omega )} \, , \label{oldmo}  
\end{equation}
where $\omega$ is the frequency of the photon,
$\rho = c^2/4\hbar^2$, $\phi$ is the mixing angle and
$\mu$ is the mass difference between the two mass eigenstates
(i.e., the mass of the additional photon if the standard one is massless).
Thus the oscillation probability decreases with the increasing photon energy.
The thermal nature of the CMB has then been established by 
COBE~\cite{cobe} and the anomaly has vanished;
from the agreement of the CMB with the blackbody radiation,
a second photon with mass $\mu \neq 0$ 
maximally mixing to the standard one has been
excluded \cite{debesmo} at the level of
\begin{equation}
\mu < 10^{-18} {\rm{eV}} \, ,
\end{equation}
to be compared with the present limit of $m_\gamma < 2 \times 10^{-16}$ eV 
on the photon mass \cite{pdg}. 
Eq.~(\ref{oldmo}) shows that, in this kind of model, 
one achieves maximum sensitivity to the mixing by studying
low-frequency radiation. 

A different model \cite{glas} has been recently motivated
by the possible existence of tiny departures from
Lorentz invariance \cite{cole}, which could explain the presence of
cosmic rays beyond the Greisen-Zatsepin-Kuzmin (GZK) 
cutoff \cite{gzk}.
An additional photon state would
experience Lorentz non-invariant  
mixing with the standard 
one, and the two eigenstates
would propagate in any direction at slightly different velocities, say, 
$c$ and $(1+\delta)\,c$.
Velocity oscillations of
photons could also result from violations of the equivalence principle in a
Lorentz invariant theory \cite{gasp}, or from the mixing with
photons in a ``shadow'' universe~\cite{sak}.

The paraphoton in \cite{glas} is sterile;
photons emitted by ordinary matter 
evolve in such a way that the non-interacting component develops with time,
and the probability for an ordinary photon to stay as such 
oscillates with time according to:
\begin{equation}
P(t) = 1-b^2\,\sin^2{(\delta \,\omega t/2)} \label{eosc} 
\end{equation}
with $\omega$ the frequency of the detected photon and 
$b^2\equiv \sin^2{(2\phi)}$, where $\phi$ is the mixing angle. 
We are concerned with the large mixing ($b\sim 1$) and small $\delta$
domain. 

\begin{figure}
\begin{center}
\includegraphics[width=0.8\linewidth]{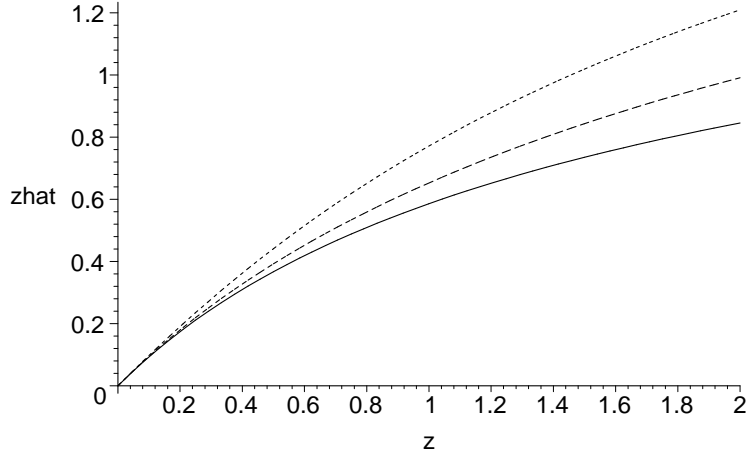}
\end{center}
\caption{The function $\hat{z}(z)$: for 
$\Omega = 1$ and $\Omega_\Lambda = 0$ (solid line); 
for $\Omega = 1$ and $\Omega_\Lambda = 0.7$ (dashed line);
and for $\Omega = 0.3$ and $\Omega_\Lambda = 0$ (dotted line).
}
\label{fig1}
\end{figure}

The extinction coefficient on light from a source at redshift $z$,
due to velocity oscillations, 
can be written as a function of $z$ as:
\begin{equation} 
P(z) =1-b^2\sin^2( \delta\,\omega \,\hat{z}/2H_0) \, , \label{eoscc} 
\end{equation}
where
\begin{equation}
\hat{z} = H_0 \int_0^{z} (1+\zeta)\,d\zeta\,(dt/d\zeta)\;, \label{erep} 
\end{equation}
$H_0=h \times 100\ {\rm  km/s}\cdot$Mpc, and the redshift-time relation
can be written:
\begin{eqnarray*}
\frac{dy}{y} & = & -H(y)\,dt=-H_0\,\left[(1-\Omega)y^2 +(\Omega-\Omega_\Lambda)y^3
+\Omega_\Lambda\right]^{1/2 }\,dt 
\end{eqnarray*}
with $y=(1+\zeta)$, 
$\Omega_\Lambda$ the cosmological constant and $\Omega_{\rm M} =
\Omega-\Omega_\Lambda$.
The function $\hat{z}(z)$ is plotted in Figure \ref{fig1} for 
$\Omega = 1$ and $\Omega_\Lambda = 0$; for $\Omega = 1$ and $\Omega_\Lambda = 0.7$;
and for $\Omega = 0.3$ and $\Omega_\Lambda = 0$ respectively.

The analysis of Ref. \cite{deangelis} improves the previous limit \cite{glas} 
by studying the signal from Type1a supernov\ae \cite{SCP}. A new limit
\begin{equation}
\delta < 2 \times 10^{-34} \, ,
\end{equation} 
for $h=0.7$ and $b=1$, is obtained.
More constraining limits could be 
reached if no distortions were observed in the
spectrum of more distant supernov\ae.

By inspecting Eq. (\ref{eosc}), one can see
that, in this kind of model, 
one achieves maximum sensitivity to the mixing by studying
high energy radiation. 

The presence of the term $\omega$ in Eq. (\ref{eoscc}) is such that the
sensitivity to $\delta$ improves further 
by making observations in the $\gamma$-ray region.
A rule-of-thumb relation on the value of $\delta$ which could have sizable
effects on the propagation from a given redshift $z$ 
of a photon of energy $E$ can be obtained by
setting to unity the argument of the sine in  Eq. (\ref{eoscc}):
\begin{equation}
\delta \sim  
\frac{3 \times 10^{-33}}{2(1-1/\sqrt{1+z})}  \, 
\left( \frac{\mathrm{1\,eV}}{E} \right) \, . \label{rot}
\end{equation}

The non-observation of distortions in the $\gamma$-ray spectrum at  
$z \sim 1$, in an energy region between 30 MeV and 20 GeV, corresponding to
the sensitivity of GLAST\cite{glast+-+} folded by the requirement of a reasonable flux,
can thus allow to exclude the region between $\delta < 10^{-40}$
and  $\delta < 10^{-43}$.

The high-energy gamma data from Mkr 501 \cite{ah99}
at a redshift $z \simeq 0.034$, if interpreted 
as in agreement with models \cite{ko99}, allow to set
a model-dependent limit
\begin{equation}
\delta < 10^{-46} 
\end{equation} 
from Eq. (\ref{rot}) assuming that there is no distortion at 
$E \sim 10$ TeV, an energy threshold which can be reached with clean data and a reasonable sensitivity in ground-based detectors like MAGIC \cite{magic} or ARGO \cite{argo}.


\begin{thebibliography}{99}

\bibitem{geo}
H.~Georgi, S.~L.~Glashow and P.~Ginsparg, Nature 306 (1983)~765

\bibitem{cobe}
D.~J.~Fixen, E.~S.~Cheng, J.~M.~Gales, J.~C.~Mather, R.~A.~Shafer and E.~L.~Wright, {\tt http://xxx.lanl.gov/astro-ph/9605054}
 
\bibitem{debesmo}
P.~de~Bernardis et al., Astrophys.~J.~284 (1984)~L21;\\
H.~P.~Nordberg and G.~F.~Smoot, \\{\tt http://xxx.lanl.gov/astro-ph/9805123}
 
\bibitem{pdg}
D.~E.~Groom et al., Eur.~Phys.~J.~C15 (2000)~1

\bibitem{glas}
S.~L.~Glashow, Phys.~Lett.~B430 (1998) 54
\bibitem{cole}
S.~Coleman and S.~L.~Glashow, Phys.~Lett.~B405 (1997)~249

\bibitem{gzk}
K.~Greisen, Phys.~Rev.~Lett.~16 (1966)~748;\\
G.~T.~Zatsepin and V.~A.~Kuzmin, JETP Lett.~4 (1966)~78
 
\bibitem{gasp}
M.~Gasperini, Phys.~Rev.~D38 (1988)~2635;\\
A.~Halprin and C.N. Leung, Phys.~Rev.~Lett.~67 (1991)~1833
 
\bibitem{sak}
B.~Holdom, Phys.~Lett.~B166 (1986)~196;\\
A.~Y.~Ignatiev, V.~A.~Kuzmin and M.~E.~Shaposnikov, Phys.~Lett.~65 (1990)~679;\\
R.~Foot, A.~Yu.~Ignatiev and R.~R.~Volkas, Phys.~Lett.~B503 (2001)~355
 
\bibitem{deangelis}
A.~De~Angelis and R.~Pain, Mod.~Phys.~Lett.~A17 (2002)~2491
\bibitem{SCP}
The Supernova Cosmology Project: S.~Perlmutter et al., Astrophys.~J.~517 (1999)~565
 
\bibitem{ah99}
F.~A.~Aharonian et al., Astron.~and Astrophys.~349 (1999)~11

\bibitem{ko99}
A.~K.~Konopelko, J.~G.~Kirk, F.~W.~Stecker and A.~Mastichiadas, Astrophys.~J.~518 (1999)~L13
 
\bibitem{glast+-+}
{\tt http://glast.gsfc.nasa.gov/}

\bibitem{magic}
R.~Stamerra, these proceedings

\bibitem{argo}
S.~Vernetto, these proceedings.

\end{thebibliography}
\end{document}